\documentclass[anonymous=false, %
               format=acmsmall, %
               review=false, %
               screen=true, %
               nonacm=true]{acmart}

\usepackage[ruled]{algorithm2e}
\usepackage{subcaption}
\usepackage{adjustbox}
\usepackage{multirow}

\graphicspath{{figures/}{../figures/}}
\usepackage{amsmath}

\DeclareMathOperator{\EX}{\mathbb{E}}

\begin{document}

\title{PPL Bench: Evaluation Framework For Probabilistic Programming Languages}

\author{Sourabh Kulkarni}
\email{skulkarni@umass.edu}
\author{Kinjal Divesh Shah}
\email{kshah97}
\author{Nimar Arora}
\email{nimarora}
\author{Xiaoyan Wang}
\email{xiaoyan0}
\author{Yucen Lily Li}
\email{yucenli}
\author{Nazanin Khosravani Tehrani}
\email{nazaninkt}
\author{Michael Tingley}
\email{tingley}
\author{David Noursi}
\email{dcalifornia}
\author{Narjes Torabi}
\email{ntorabi}
\author{Sepehr Akhavan Masouleh}
\email{sepehrakhavan}
\author{Eric Lippert}
\email{ericlippert}
\author{Erik Meijer}
\email{erikm}
\affiliation{%
  \institution{Facebook Inc}
  \department{Probability}
  \streetaddress{1 Hacker Way}
  \city{Menlo Park}
  \state{CA}
  \postcode{94025}
  \country{USA (@fb.com)}
}

\begin{abstract}
  We introduce PPL Bench, a new benchmark for evaluating Probabilistic Programming Languages (PPLs) on a variety of statistical models.
The benchmark includes data generation and evaluation code for a number of models as well as implementations in some common PPLs.
All of the benchmark code and PPL implementations are available on Github.
We welcome contributions of new models and PPLs and as well as improvements in existing PPL implementations.
The purpose of the benchmark is two-fold.
First, we want researchers as well as conference reviewers to be able to evaluate improvements in PPLs in a standardized setting.
Second, we want end users to be able to pick the PPL that is most suited for their modeling application.
In particular, we are interested in evaluating the accuracy and speed of convergence of the inferred posterior.
Each PPL only needs to provide posterior samples given a model and observation data.
The framework automatically computes and plots growth in predictive log-likelihood on held out data in addition to reporting other common metrics such as effective sample size and $\hat{r}$.

\end{abstract}

\maketitle

\section{Introduction}
Probabilistic Programming Languages \citep{ghahramani2015probabilistic} allow statisticians to write probability models in a formal language.
These languages usually include a builtin inference algorithm that allows practitioners to rapidly prototype and deploy new models.
More formally, given a model $P(X, Z)$ defined over random variables $X$ and $Z$ and given data $X=x$, the inference problem is to compute $P(Z|X=x)$.
Another variant of this inference problem is to compute the expected value of some functional $f$ over this posterior, $\EX[f(Z)|X=x]$.
Most languages provide some version of Markov Chain Monte Carlo \citep{brooks2011handbook} inference.
MCMC is a very flexible inference technique that can conveniently represent high dimensional posterior distributions with samples $Z_1, \ldots, Z_m$ that collectively approach the posterior asymptotically with $m$.
Other inference techniques that are often builtin include Variational Inference \citep{wainwright2008graphical}, which is usually asymptotically inexact, and Importance Sampling \citep{glynn1989importance}, which is only applicable to computing an expectation.

In roughly the last two decades there has been an explosive growth in the number of PPLs.
Starting with BUGS~\citep{spiegelhalter1996bugs}, iBAL~\citep{pfeffer2001ibal}, BLOG~\citep{milch2005blog}, and Church~\citep{goodman2008church} in the first decade, and followed by Stan~\citep{carpenter2017stan}, Venture~\citep{mansinghka2014venture}, Gen~\citep{cusumano2018design}, Turing~\citep{ge2018turing}, to name some, in the next decade.
Some of these PPLs restrict the range of models they can handle whereas others are universal languages, i.e, they support any computable probability distribution.
There is a tradeoff between being a universal language and performing fast inference because if a language only supports select probability distributions, its inference methods can optimize for that.
Different PPLs can be better suited for different use-cases and hence having a benchmarking framework can prove to be extremely useful.
We introduce PPL Bench as a benchmarking framework for evaluating different PPLs on models.
PPL Bench has already been used to benchmark a new inference method, Newtonian Monte Carlo~\citep{arora2019nmc}.
In the rest of this paper, we provide an overview of PPL Bench\footnote{Open Source on Github: https://github.com/facebookresearch/pplbench} and describe each of the initial statistical models that are included before concluding.

\section{System Overview}
\begin{figure}[!t]
  \centering
  \includegraphics[width=0.6\linewidth]{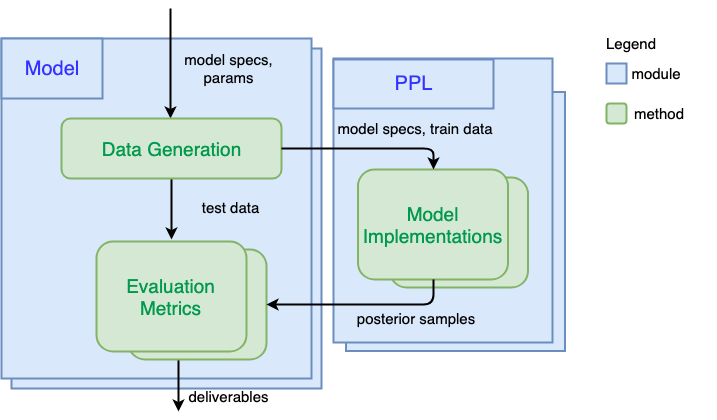}
  \caption{PPL Bench system overview}
  \label{fig:overview}
  \Description{System Overview}
\end{figure}
PPL Bench is implemented in Python 3~\citep{Python}.
It is designed to be modular, which makes it very easy to add models and PPL implementations to the framework.
The typical PPL Bench flow is as follows:
\begin{itemize}
\item \textbf{Model Instantiation and Data Generation}: All the models in PPL Bench have various parameters that need to be specified.
A model with all it's parameters set to certain values is referred to as a model instance.
We establish a model $P_\theta(X,Z)$ with ground-truth parameter distributions($\theta \sim P_\phi$) where $\phi$ is the hyperprior (a prior distribution from which parameters are sampled).
In some models theta is sampled from $\phi$ while in others it can be specified by users. PPL Bench is designed to accept parameters at both the PPL level, as well as the model level.
Therefore, any parameters such as number of samples, number of warmup samples, inference method and so on
can be configured with ease. Similarly, any model-specific hyperparameters can also be passed as input through a JSON file.
We can sample parameter values form their distributions, which will instantiate the model:
$$
Z_1 \sim P_\theta(Z)
$$
We then simulate train and test data as follows:
$$
X_{train} \stackrel{iid}{\sim} P_\theta(X|Z=Z_1)
$$
$$
X_{test} \stackrel{iid}{\sim} P_\theta(X|Z=Z_1)
$$
Note that this process of data generation is performed independent of any PPL.
\item \textbf{PPL Implementation and Posterior Sampling}: The training data is passed to various PPL implementations which learn a model $P_\theta(X = X_1,Z)$;
The output of this learning process is obtained in the form of $n$ sets of posterior samples of parameters (one for each sampling step).
$$
Z^*_{1...n} \sim P_\theta(Z | X = X_{train})
$$

The sampling is restricted to a single chain and any form of multi-threading is disabled. However, we can run each benchmark multiple times, and we treat each trial as a chain. This allows us to calculate metrics such as $r_{hat}$.
\item \textbf{Evaluation}: Using the test data $X_{test}$, posterior samples $Z^*_{1...n}$ and timing information, the PPL implementations are evaluated on the following evaluation metrics:\newline
\textbf{Plot of the Predictive Log Likelihood w.r.t samples for each implemented PPL}:
PPL Bench automatically generates a plot of predictive log likelihood against samples. This provides a visual representation
of how fast different PPLs converge to their final predictive log likelihood value.
This metric has been previously used by \cite{kucukelbir2017automatic} to compare convergence of different implementations of statistical models.
$$
 \text{Predictive Log Likelihood}(n) = \log \left( \frac{1}{n}\sum_{i=1}^{n}(P(X_{test}|Z=Z^*_{i})) \right)
$$

After computing the predictive log likelihood w.r.t samples over each trial run, the min, max and mean values over samples is obtained.
These are plotted on a graph for each PPL implementation so their convergence behavior can be visualized.
This plot is designed to capture both the relative performance and accuracy of these implementations. \newline
\textbf{Gelman-Rubin convergence statistic $r_{hat}$}: We report $r_{hat}$ for each PPL as a measure of convergence by using samples generated across trials.
We report this quantity for all queried variables.
\newline
\textbf{Effective sample size $n_{eff}$}: The samples generated after convergence should ideally be independent of each other.
In reality, there is non-zero correlation between generated samples; any positive correlation reduces the number of effective samples generated.
Samples are combined across trials and the $n_{eff}$ as well as $n_{eff}/s$ of each queried variable in each PPL implementation is computed. \newline
\textbf{Inference time}: Runtime is an important consideration for any practical use of probabilistic models.
The time taken to perform inference in each trial of a PPL is recorded.
\end{itemize}

The generated posterior samples and the benchmark configurations such as number of samples, ground-truth model parameters etc. are stored
along with the plot and evaluation metrics listed above for each run.
PPL Bench also allows for model-specific evaluation metrics to be added.
Using the PPL Bench framework, we establish four models to evaluate performance of PPLs:

\begin{enumerate}

\item Bayesian Logistic Regression model\cite{vanErp2013}
\item Robust Regression model with Student-T errors\cite{gelman2013bayesian}
\item Latent Keyphrase Index(LAKI) model, a Noisy-Or Topic Model\cite{liu2016representing}
\item Crowdsourced Annotation model, estimating true label of an item given a set of labels assigned by imperfect labelers\cite{passonneau2014benefits}
\end{enumerate}
These models are chosen to cover a wide area of use cases for probabilistic modeling.
Each of these models is implemented in all or a subset of these PPLs: PyMC3~\citep{Salvatier2016}, JAGS~\citep{plummer2003jags}, Stan~\citep{carpenter2017stan} and BeanMachine~\citep{arora2020bm}.
The model and PPL implementation portfolio is expected to be expanded with additional open-source contributions.
Subsequent sections describe these models and implementations in detail along with analysis of observed results.

\section{Bayesian Logistic Regression Model}
\begin{figure}[h]
  \centering
  \begin{subfigure}{.5\linewidth}
    \centering
    \includegraphics[width=0.9\linewidth]{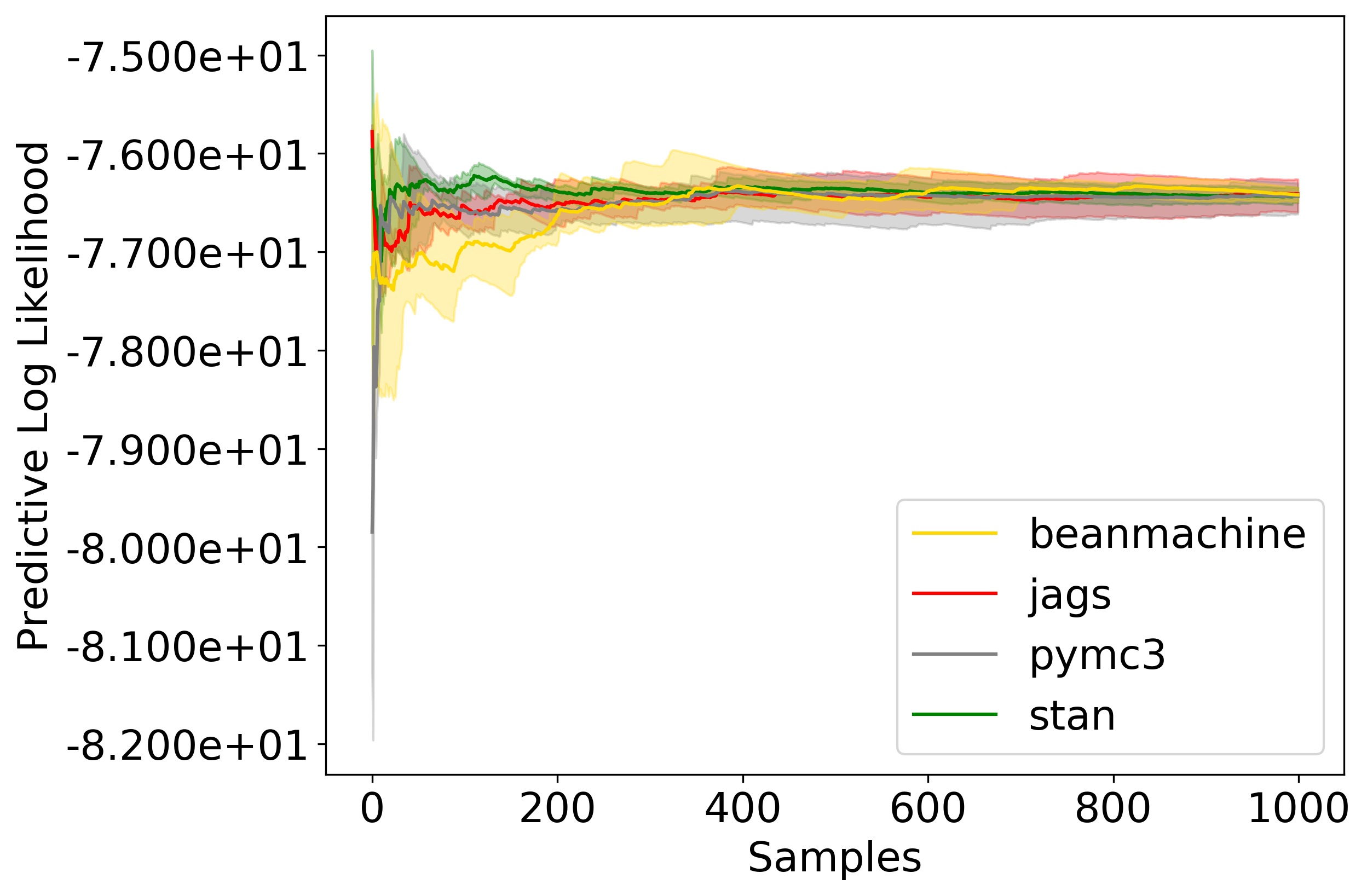}
    \caption{N = 20K, K = 10}
    \label{fig:blr1}
  \end{subfigure}%
  \begin{subfigure}{.5\linewidth}
    \centering
    \includegraphics[width=.9\linewidth]{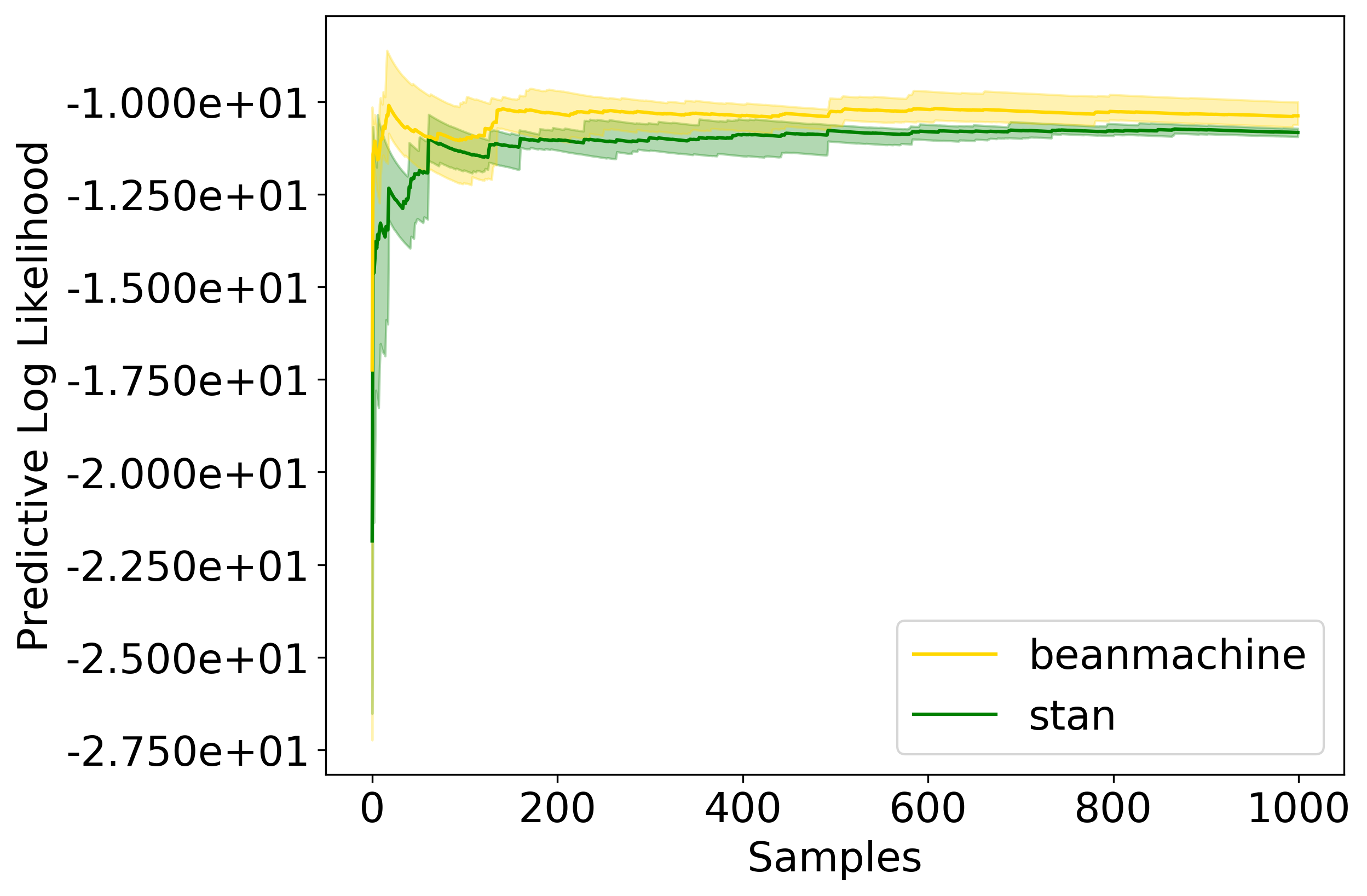}
    \caption{N = 200K, K = 10}
    \label{fig:blr2}
  \end{subfigure}
  \caption{Predictive log likelihood against samples for Bayesian Logistic Regression}
  \label{fig:blr}
\end{figure}

Logistic Regression is the one of the simplest models we can benchmark PPLs on.
Because its posterior is log-concave, it is easy for PPLs to converge to the
true posterior.
For model definition, see Appendix \ref{model:blr}.
Here, $N$ refers to number of data points and $K$ refers to the number of covariates.
Figure~\ref{fig:blr} shows the predictive log likelihood for $N = 20k$ and $N = 200k$.
We use half of the data for inference and the other half for evaluation.
We run $1000$ warm-up iterations for each of the PPLs and plot the samples from
the next $1000$ iterations. We can see from Figure~\ref{fig:blr} that
all the predictive log likelihood (PLL) of different PPLs converge to around the same
value. Thus, in addition to convergence, the plot could also be used to check the accuracy of the model.

\section{Robust Regression Model}
Bayesian logistic regression with Gaussian errors, like ordinary least-square regression,
is quite sensitive to outliers \cite{rousseeuw2005robust}.
To increase outlier robustness, a Bayesian regression model with Student-T errors is more effective \cite{gelman2013bayesian}. For model definition, see Appendix \ref{model:robust}.

\begin{table}
  \begin{minipage}{.45\linewidth}
    \centering
    \begin{adjustbox}{width=\linewidth,center}
      \begin{tabular}{cccccc}
        \toprule
        \multirow{2}{*}{PPL} & \multirow{2}{*}{N} & \multirow{2}{*}{Time(s)} & \multicolumn{3}{c}{$n_{eff}$/s}                   \\
                             &                    &                          & min                             & median & max    \\

        \midrule
        BeanMachine          & 20K                & 298.64                   & 21.74                           & 232.85 & 416.98 \\
        Stan                 & 20K                & 47.06                    & 21.62                           & 25.78  & 108.63 \\
        Jags                 & 20K                & 491.22                   & 0.17                            & 0.18   & 4.04   \\
        PyMC3                & 20K                & 48.37                    & 23.65                           & 27.10  & 67.27  \\
        \hline
        BeanMachine          & 200K               & 1167.13                  & 0.003                           & 0.004  & 0.01   \\
        Stan                 & 200K               & 10119.02                 & 0.06                            & 0.06   & 0.34   \\
        \bottomrule
      \end{tabular}
    \end{adjustbox}
    \captionof{table}{Runtime and $n_{eff}$/s for Bayesian Logistic Regression.}
    \label{table:blr}
  \end{minipage}%
  \hspace{3mm}
  \begin{minipage}{.45\linewidth}
    \begin{adjustbox}{width=\linewidth,center}
      \begin{tabular}{cccccc}
        \toprule
        \multirow{2}{*}{PPL} & \multirow{2}{*}{N} & \multirow{2}{*}{Time(s)} & \multicolumn{3}{c}{$n_{eff}$/s}                   \\
                             &                    &                          & min                             & median & max    \\
        \midrule
        BeanMachine          & 20K                & 146.79                   & 3.11                            & 8.48   & 41.85  \\
        Stan                 & 20K                & 35.21                    & 87.71                           & 114.75 & 361.16 \\
        Jags                 & 20K                & 742.40                   & 0.59                            & 1.40   & 2.87   \\
        PyMC3                & 20K                & 45.78                    & 92.50                           & 99.14  & 137.45 \\
        \hline
        BeanMachine          & 200K               & 250.41                   & 1.87                            & 4.00   & 10.33  \\
        Stan                 & 200K               & 140.58                   & 0.77                            & 1.52   & 2.40   \\
        \bottomrule
      \end{tabular}
    \end{adjustbox}
    \captionof{table}{Runtime and $n_{eff}$/s for Robust Regression}
    \label{table:robust}
  \end{minipage}
\end{table}

\begin{figure}[h]
  \centering
  \begin{subfigure}{.5\linewidth}
    \centering
    \includegraphics[width=0.9\linewidth]{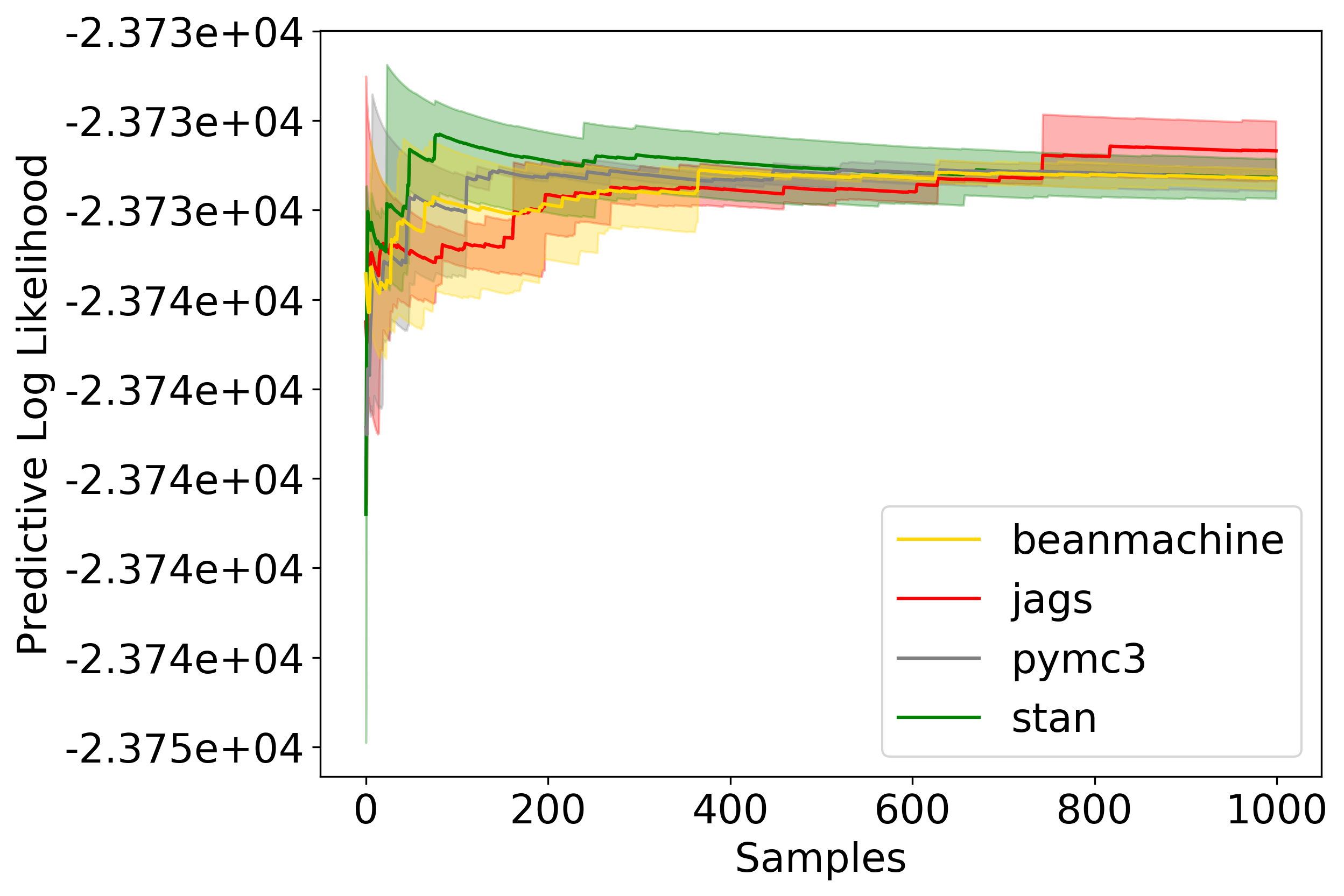}
    \caption{N = 20K, K = 10}
    \label{fig:robust1}
  \end{subfigure}%
  \begin{subfigure}{.5\linewidth}
    \centering
    \includegraphics[width=.9\linewidth]{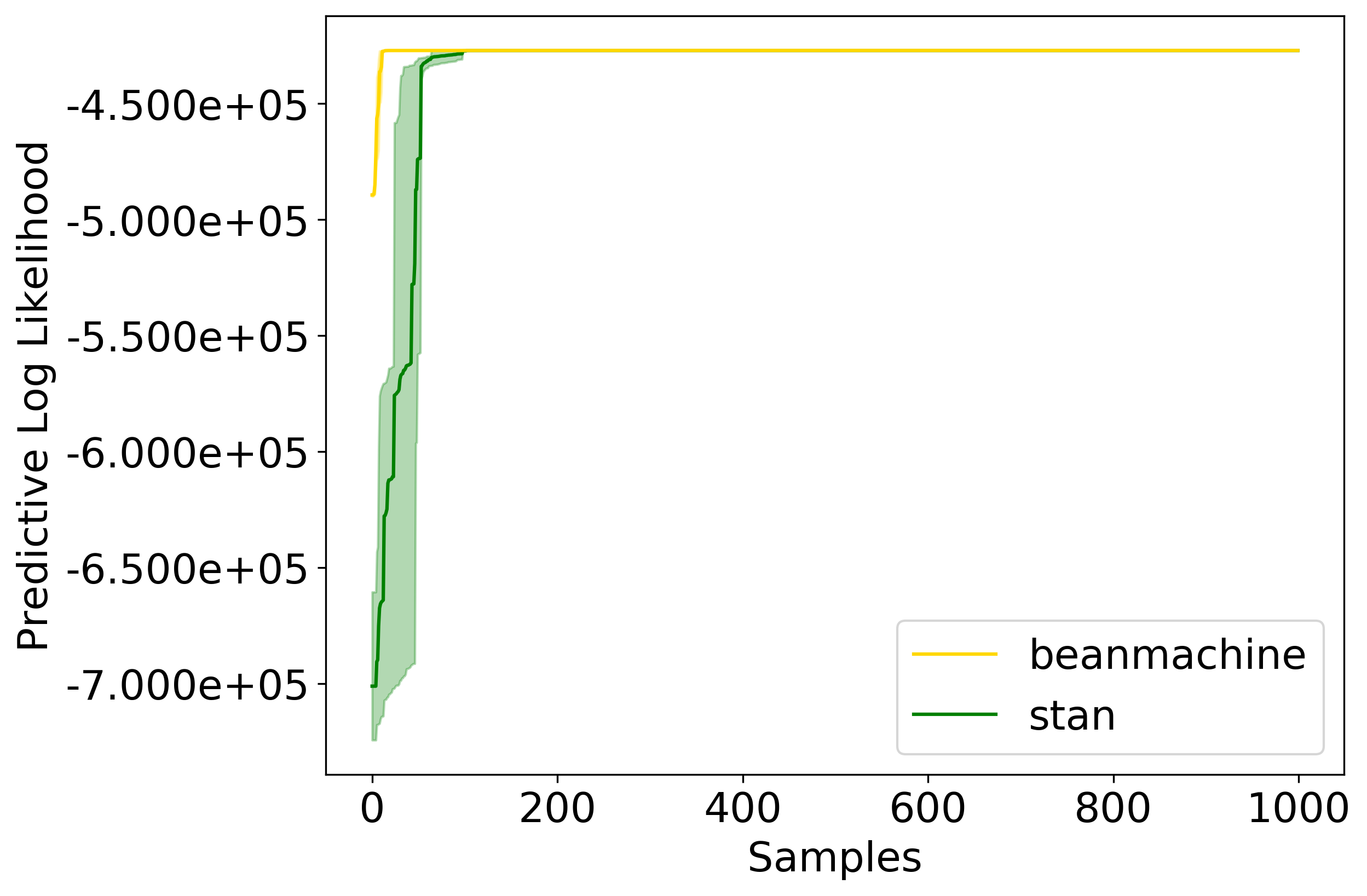}
    \caption{N = 200K, K = 10 (warm up samples included)}
    \label{fig:robust2}
  \end{subfigure}
  \caption{Predictive log likelihood against samples for Robust Regression}
  \label{fig:robust}
\end{figure}

In Figure~\ref{fig:robust2},
we include warmup samples in the plot. By doing so, we can also gain insights
into how quickly an algorithm adapts. We can see that Bean Machine adapts quicker than Stan.
Moreover, depending on the use-case, some metrics might be more
informative to the end-users than others. For example, we can see from
Table~\ref{table:robust} that Stan is faster than BeanMachine,
but BeanMachine generates more independent samples per time as $n_{eff}$/s is
higher.

\section{Noisy-Or Topic Model}
Inferring topics from the words in a document is a crucial task in many natural language processing tasks.
The noisy-or topic model~\citep{liu2016representing} is one such example. It is a Bayesian network consisting of two types of nodes; one type are the domain key-phrases (topics) that are latent while others are the content units (words) which are observed.
Words have only topics as their parents. To capture hierarchical nature of topics, they can have other topics as parents. For model definition, see Appendix \ref{model:noisyor}.

For the experiment, we first sample the model structure. This consists of determining the children of each node in the graph, and the corresponding weight associated with the edge.
Next, the key-phrase nodes are sampled once; and two instances of content units are sampled keeping the key-phrase nodes fixed. One instance is passed to PPL implementation for inference, while the other is used to compute posterior predictive of the obtained samples.

\subsection{Stan Implementation}
The model requires support for discrete latent variables, which Stan does not have.
As a workaround, we did a reparameterization; the discrete latent variables with values "True" and "False" which denote whether a node is activated or not are instead represented as 1-hot encoded vectors\cite{maddison2016concrete}. Each element of the encoded vector is now a real number substituting a boolean variable.
"True" is represented by $[1, 0]$ and "False" by $[0, 1]$ in the 1-hot encoding; a relaxed representation of a true value, for example, might look like [0.9999, 0.0001].
The detailed reparameterization process is as follows:
We encode the probabilities of a node in a parameter vector $\alpha = [\alpha_{\text{true}}, \alpha_{\text{false}}]$, then choose a temperature parameter $\tau = 0.1$. For each key-phrase, we assign an intermediate random variable $X_j$:

\begin{table}[h]
  \begin{minipage}[b]{0.4\linewidth}
    \includegraphics[width=1.0\columnwidth]{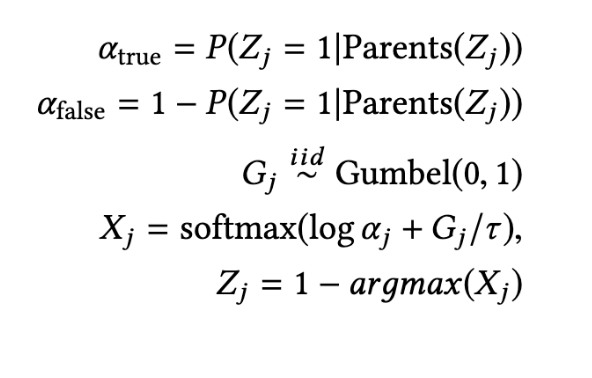}
  \end{minipage}\hfill
  \begin{minipage}[b]{0.6\linewidth}
    \begin{adjustbox}{width=1.0\linewidth,center}
      \begin{tabular}{ccccccc}
        \toprule
        \multirow{2}{*}{PPL} & \multirow{2}{*}{Words} & \multirow{2}{*}{Topics} & \multirow{2}{*}{Time(s)} & \multicolumn{3}{c}{$n_{eff}$/s}                       \\
                             &                        &                         &                          & min                             & median   & max      \\
        \midrule
        Stan                 & 300                    & 30                      & 37.58                    & 17.50                           & 79.83    & 80.39    \\
        Jags                 & 300                    & 30                      & 0.17                     & 15028.96                        & 17350.35 & 18032.36 \\
        PyMC3                & 300                    & 30                      & 38.05                    & 39.50                           & 78.83    & 79.33    \\
        \hline
        Stan                 & 3000                   & 100                     & 438.74                   & 2.27                            & 6.84     & 7.00     \\
        Jags                 & 3000                   & 100                     & 0.68                     & 3773.66                         & 4403.28  & 4504.26  \\
        PyMC3                & 3000                   & 100                     & 298.79                   & 3.96                            & 10.04    & 14.72    \\
        \bottomrule
      \end{tabular}
    \end{adjustbox}
    \captionof{table}{Runtime and $n_{eff}$/s for Noisy-Or Topic Model.}
    \label{table:noisy_or}
  \end{minipage}
\end{table}

\begin{figure}[h]
  \centering
  \begin{subfigure}{.5\linewidth}
    \centering
    \includegraphics[width=0.9\linewidth]{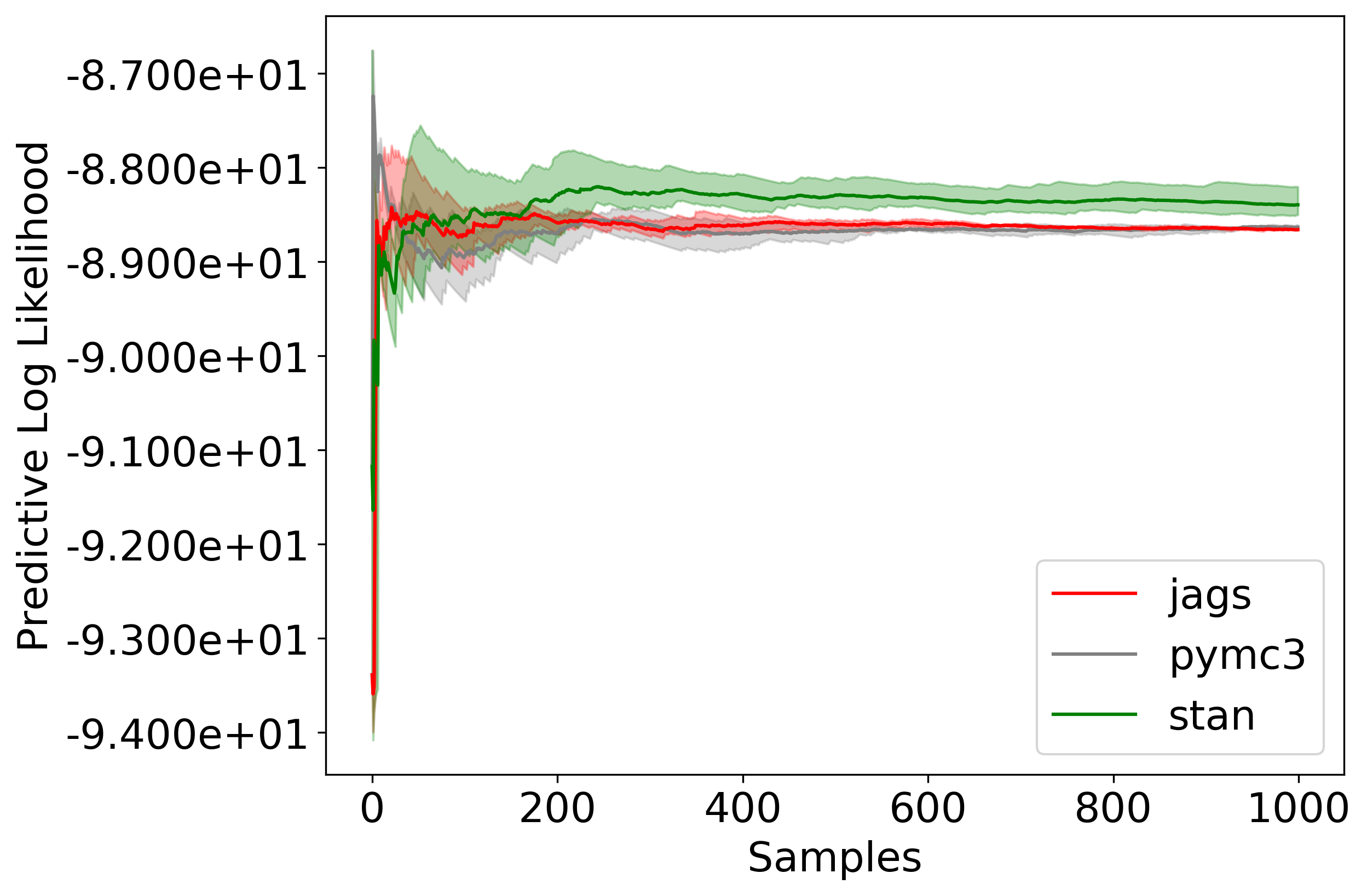}
    \caption{30 Topics, 300 Words}
    \label{fig:noisyor1}
  \end{subfigure}%
  \begin{subfigure}{.5\linewidth}
    \centering
    \includegraphics[width=.9\linewidth]{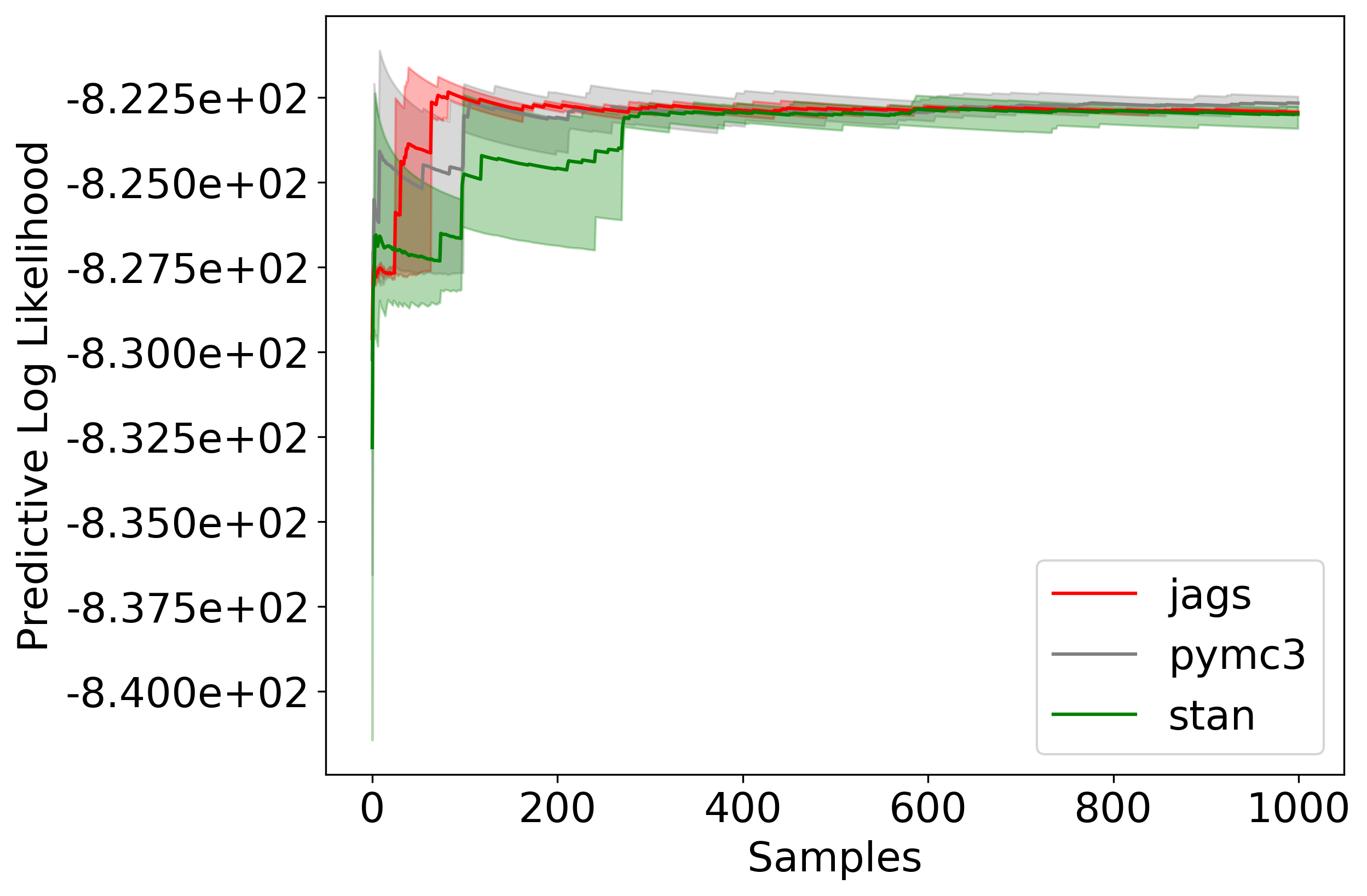}
    \caption{100 Topics, 3000 Words}
    \label{fig:noisyor2}
  \end{subfigure}
  \caption{Predictive log likelihood against samples for Noisy-Or Topic Model}
  \label{fig:noisyor}
\end{figure}

\vspace{-8mm}
From Figure~\ref{fig:noisyor}, we can see that all PPLs converge to the same predictive log likelihood.
We see that JAGS is extremely fast at this problem because it is written in C++ while PyMC3 is Python-based.
We suspect that Stan is slower then PyMC3 and JAGS because of the reparameterization which introduces twice
as many variables per node.

\section{Crowdsourced Annotation Model}
There exist several applications where the complexity and volume of the task requires crowdsourcing to a large set of human labelers.
Inferring the true label of an item from the labels assigned by several labelers is facilitated by this crowdsourced annotation model\cite{passonneau2014benefits}.
The model takes into consideration an unknown prevalence of label categories, an unknown per-item category, and an unknown confusion matrix per-labeler. For model definition, see \ref{model:crowd_sourced}

\subsection{Results}

\begin{center}
  \begin{minipage}[b]{0.45\linewidth}
    \includegraphics[width=1.0\columnwidth]{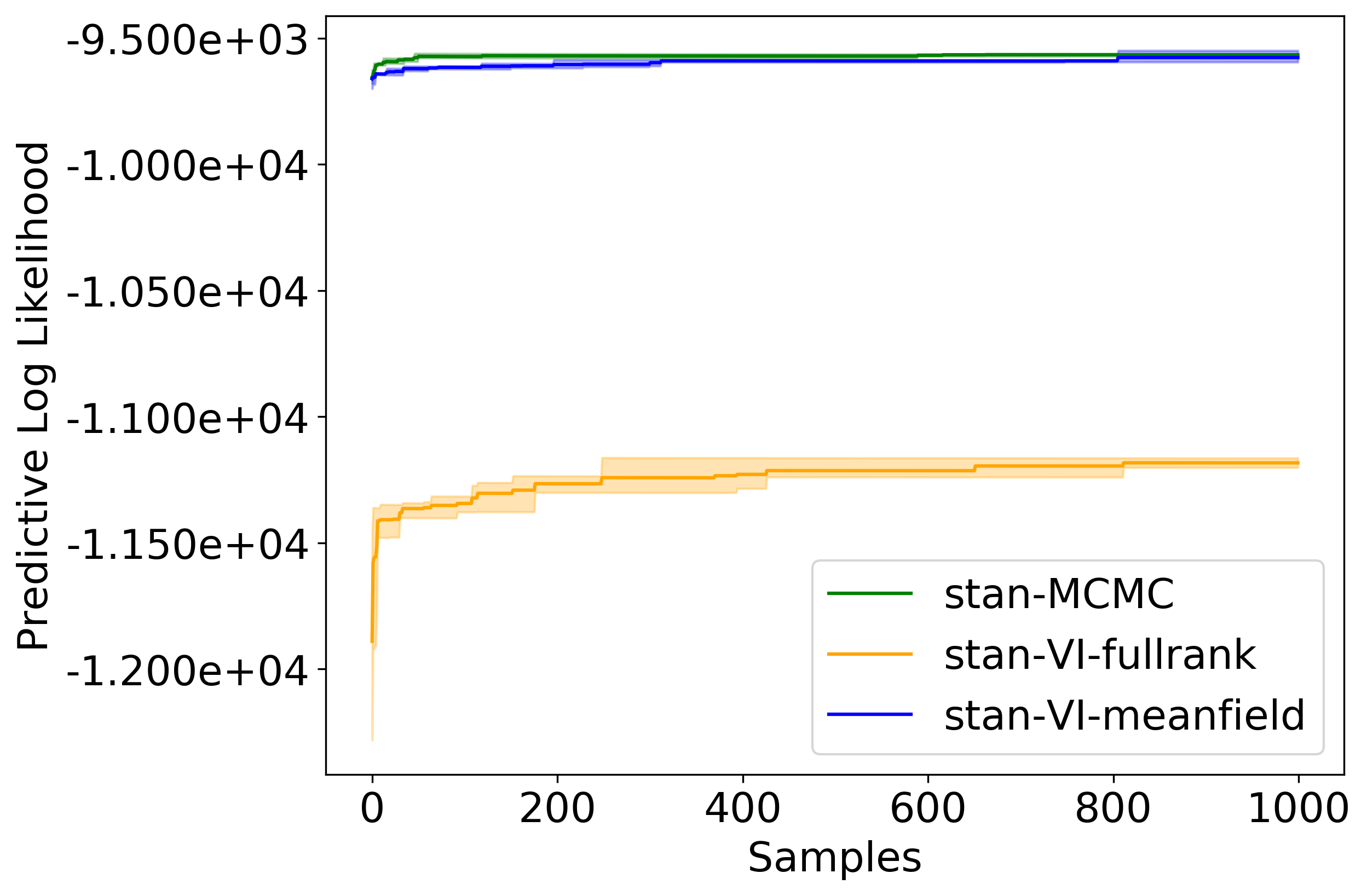}
    \captionof{figure}{Predictive log likelihood against samples for Crowd Sourced Annotation Model}
    \label{fig:annotation}
  \end{minipage}\hfill
  \hspace{3mm}
  \begin{minipage}[b]{0.45\linewidth}
    \begin{adjustbox}{width=1.0\linewidth}
      \begin{tabular}{ccccc}
        \toprule
        \multirow{2}{*}{PPL} & \multirow{2}{*}{Time(s)} & \multicolumn{3}{c}{$n_{eff}$/s}                   \\
                             &                          & min                             & median & max    \\
        \midrule
        Stan-MCMC            & 484.59                   & 2.14                            & 4.22   & 10.45  \\
        Stan-VI (meanfield)  & 15.23                    & 0.48                            & 7.86   & 216.35 \\
        Stan-VI (fullrank)   & 35.79                    & 0.12                            & 0.20   & 84.16  \\
        \bottomrule
        \vspace{1cm}
      \end{tabular}
    \end{adjustbox}
    \captionof{table}{Runtime and $n_{eff}$/s for Crowd Sourced Annotation Model with items=10K and \mbox{labelers=100}}
    \label{table:annotation}
  \end{minipage}
\end{center}

PPL Bench can also benchmark different inference techniques within a PPL.
Here, we benchmark Stan's NUTS against Stan's VI methods, using both meanfield as well as fullrank algorithm.
We notice that VI using meanfield algorithm is indistinguishable from NUTS.
However, VI using fullrank converges to a slightly different predictive log likelihood.

\section{Conclusion}
We have provided a modular benchmarking framework for evaluating PPLs.
In its current form, this is a very minimal benchmarking suite, but we hope that it will become more diverse with community contributions.
Our work to standardize the evaluation of inference should aid scientific development in this field, and our focus on popular statistical models should encourage overall industrial deployments of PPLs.

\bibliographystyle{acm-reference-format}
\bibliography{references}

\section{Appendix A - Statistical Model Definitions}
\subsection{Bayesian Logistic Regression}
\label{model:blr}
\begin{align*}
\alpha &\sim \text{Normal}(0, 10, \text{size}=1),\\
\beta &\sim \text{Normal}(0, 2.5, \text{size}=K),\\
X_i &\sim \text{Normal}(0,\ 10,\ \text{size}=K) \quad \forall i \in 1\ldots N\\
\mu_i &= \alpha + {X_i}^T \beta \quad \forall i \in 1..N\\
Y_i &\sim \text{Bernoulli}(\text{logit}=\mu_i) \quad \forall i \in 1..N.
\end{align*}

\subsection{Robust Regression}
\label{model:robust}
\begin{align*}
\nu &\sim \text{Gamma}(2,\ 10)\\
\sigma &\sim \text{Exponential}(1.0)\\
\alpha &\sim \text{Normal}(0,10)\\
\beta &\sim \text{Normal}(0,\ 2.5,\ \text{size}=K)\\
X_i &\sim \text{Normal}(0,\ 10,\ \text{size}=K) \quad \forall i \in 1\ldots N\\
\mu_i &= \alpha + \beta^T X_i  \quad \forall i \in 1\ldots N\\
Y_i &\sim \text{Student-T}(\nu,\ \mu_i,\ \sigma) \quad \forall i \in 1\ldots N\\
\end{align*}

\subsection{Noisy-OR Topic Model}
\label{model:noisyor}
Let $Z$ be the the set of all nodes in the network.
Each model has a leak node $O$ which is the parent of all other nodes.
The set of keyphrases is denoted by $K$ and the set of content units is represented as $T$.
\begin{align*}
|\text{Children}(K_i)| \sim \text{Poisson}(\lambda = 3) \quad \forall i \in 1\ldots K \\
\text{W}_{oj} \sim \text{Exponential}(0.1) \quad \forall j \in 1\ldots Z\\
\text{W}_{ij} \sim \text{Exponential}(1.0) \quad \forall i,j \text{ where i} \in \text{Parents}(Z_j)\\
P(Z_j = 1|\text{Parents}(Z_j)) = 1 - \exp(-W_{oj} -\sum_{i}(W_{ij} * Z_i)) \\
\text{Z}_j \sim \text{Bernoulli}(P(Z_j = 1|\text{Parents}(Z_j))) \\
\end{align*}

\subsection{Crowd Sourced Annotation Model}
\label{model:crowd_sourced}

There are $N$ items, $K$ labelers, and each item could be one of $C$ categories.
Each item $i$ is labeled by a set $J_i$ of labelers. Such that the size of $J_i$ is sampled randomly, and each labeler in $J_i$ is drawn uniformly without replacement from the set of all labelers.
$z_i$ is the true label for item $i$ and $y_{ij}$ is the label provided to item $i$ by labeler $j$.
Each labeler $l$ has a confusion matrix $\theta_l$ such that $\theta_{lmn}$ is the probability that an item with true class $m$ is labeled $n$ by $l$.
\begin{align*}
\pi &\sim \text{Dirichlet}\left(\frac{1}{C}, \ldots, \frac{1}{C}\right)\\
z_i &\sim \text{Categorical}(\pi)\quad \forall i \in 1\ldots N\\
\theta_{lm} &\sim \text{Dirichlet}(\alpha_m)\quad \forall l \in 1\ldots K,\ m \in 1 \ldots C\\
|J_i| &\sim \text{Poisson}(J_\text{loc})\\
l \in J_i &\sim \text{Uniform}(1\ldots K) \quad \text{without replacement}\\
y_{il} &\sim \text{Categorical}(\theta_{l z_i})\quad \forall l \in J_i\\
\end{align*}
Here $\alpha_m \in {\mathbb{R}^+}^C$.
We set $\alpha_{mn} = \gamma\cdot \rho$ if $m=n$ and $\alpha_{mn} = \gamma \cdot (1 - \rho)\cdot \frac{1}{C-1}$ if $m \ne n$.
Where $\gamma$ is the concentration and $\rho$ is the {\em a-priori} correctness of the labelers.
In this model, $Y_{il}$ and $J_i$ are observed.
In our experiments, we fixed $C=3$, $J_\text{loc}=2.5$, $\gamma=10$, and $\rho=0.5$.

\end{document}